\documentclass{iopart}
\newcommand{\bm}[1]{ \mbox{\boldmath $#1$}  }
\usepackage{graphicx}
\usepackage{epsfig}

\begin{document}

\title{Triple charged-particle decays of resonances illustrated by
$^{12}$C-states}

\author{R~\'Alvarez-Rodr\'{\i}guez$^1$, E~Garrido$^2$, A~S~Jensen$^1$,
D~V~Fedorov$^1$ and H~O~U~Fynbo$^1$}

\address{ $^1$ Institut for Fysik og Astronomi, Aarhus Universitet DK-8000 
Aarhus C, Denmark\\  
$^2$ Instituto de Estructura de la Materia, CSIC E-28006 Madrid, Spain}

\ead{raquel@phys.au.dk}

\begin{abstract}
The hyperspherical adiabatic expansion is combined with complex
scaling and used to calculate the energy distributions of the
particles arising from three-body decaying low-lying $^{12}$C resonances. The
large distance continuum properties of the wavefunctions are crucial
and must be accurately calculated. The substantial changes from small
to large distances determine the decay mechanisms. We illustrate by
computing the energy distributions from decays of the $1^{-}$, $2^-$ and
$4^{-}$-resonances in $^{12}$C. These states are dominated by
sequential ($1^-$), through the $^{8}$Be ground state, and direct ($2^-$, 
$4^-$) decays.
\end{abstract}


\submitto{\jpg}

%\maketitle

\section{Introduction.}

The importance of the triple-$\alpha$ process is well known in nuclear
astrophysics. The process leads from three free $\alpha$-particles
via a low-lying resonance into the ground state of $^{12}$C. The
inverse of the crucial part of this process is the decay of a
resonance of $^{12}$C into the continuum of three
$\alpha$-particles, perhaps via intermediate states of $^{8}$Be \cite{fynbo05}.
The principle of detailed balance relates direct and inverse processes.
Similar processes involving three charged particles occur at
specific waiting points for the $rp$-process \cite{gri05}.

Moreover, the breakup of a physical system into a three-body continuum with
Coulomb interaction is not yet a well understood problem of few body
physics, although it has been studied over many years. In the three-body
final state the asymptotics are determined by the dynamics of the
breakup process itself. The study of these processes would help us to
find out about the inverse process (triple-$\alpha$ and 2p capture) and to study
to what extent the three particles of the final state may be present
as a cluster structure in the many-body initial state. The main
difficulty lies in constructing correct asymptotic wave functions when
both two- and three-body structures are present.

Several published experimental results strongly suggest that the decay
mechanism varies from sequential decay, via an intermediate
quasi-stable state, to direct decay into the three-body continuum
\cite{fynbo05}.  The corresponding energy distributions exhibit
completely different shapes. Reliable model computations
reproducing the measured energy distributions are not available.  This
may be understandable due to several indispensable difficult
requirements, but it is nevertheless very unfortunate because
quantitatively accurate models are needed to extract and understand
the underlying physics contained in the increasing amount of
experimental high-quality data sets.

In the present paper we exploit a reliable recently developed technique to compute the
large distance properties of a many body resonance that decays into three particles.
We shall use the hyperspherical adiabatic expansion method of the Faddeev
equations combined with complex scaling of the coordinates \cite{nie01}.
The method first provides an accurate computation of the resonance
properties, which determines the energy distribution between the three 
particles after the decay.

\section{Theoretical framework.}

We describe $^{12}$C as a three-$\alpha$ cluster system, since triple-$\alpha$
decay is the only open non-electromagnetic decay channel for this nucleus.
We use the hyperspherical adiabatic expansion method combined with complex 
scaling to compute the resonance wavefunctions \cite{nie01}. The 
so-called hyperradius $\rho$ is the most important of the coordinates, and 
is defined as
\begin{eqnarray}  
  m_N  \rho^{2} =  \frac{m_{\alpha}}{3} \sum^{3}_{i<j}
 \left(\bm{r}_{i}-\bm{r}_j\right)^{2} = 
  m_{\alpha}  \sum^{3}_{i=1} 
 \left(\bm{r}_{i}-\bm{R}\right)^{2}  \label{e120} \;,
\end{eqnarray}
for three identical particles of mass $m_\alpha = 4m_N$, where 
we choose $m_N$ to be equal to the nucleon mass, 
$\bm{r}_{i}$ is the coordinate of the i-th particle and $\bm{R}$ 
is the centre-of-mass coordinate. 
 
We first choose the two-body interaction to reproduce the low-energy
two-body scattering properties \cite{ali66}.
To adjust the energy position of the many-body resonance we add then a three-body 
potential, whose range corresponds 
to three touching $\alpha$-particles. These potentials are chosen 
independently for each $J^\pi$.

The total wave function is expanded on the angular 
eigenfunctions $\Phi_{nJM} (\rho,\Omega)$ obtained as solutions to the Faddeev 
equations for fixed $\rho$,
\begin{equation}
\Psi^{JM} = \frac{1}{\rho^{5/2}}\sum_n f_n (\rho) \Phi_{nJM} (\rho,\Omega)\;,
\end{equation}
where the radial coefficients $f_n (\rho)$ are obtained from
the coupled set of radial equations. The effective adiabatic potentials
are the eigenvalues of the angular part.

The three-body potential is used to adjust the small-distance part of the effective 
potential in order to reproduce the correct resonance energies; at intermediate
distances the potential has a barrier that determines the resonance width;
and at large distances the resonance wave functions contain information about
distributions of relative energies.

The large-distance asymptotic behaviour of the decaying resonance wave function determines
the observable energy distribution. This energy 
distribution can be computed in coordinate space, except for a 
phase-space factor, as the integral of the absolute square of the 
total wave function for a large value of the hyperradius \cite{gar07a}. We shall 
explore the conjecture that the final state can be described entirely
within the present cluster model.

\section{Results.}

\begin{figure}
\vspace*{-2.51cm}
\epsfig{file=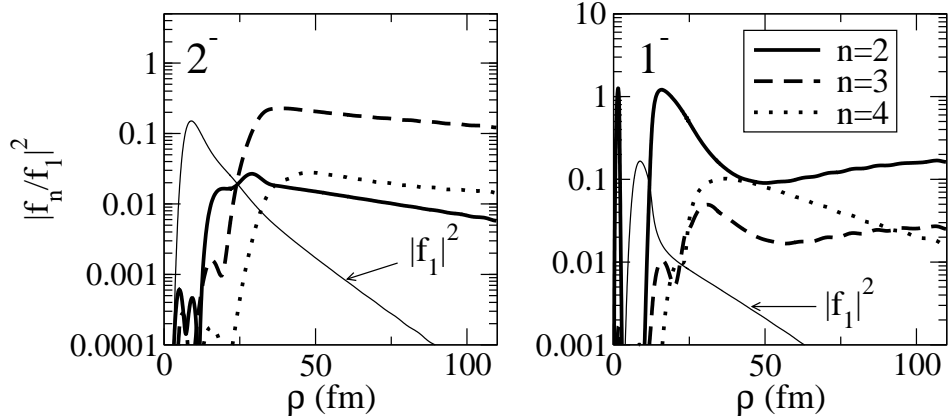,angle=-90,scale=0.51}
\vspace*{-2.3cm}
\caption{Ratios between the four lowest adiabatic radial wave
functions and the small-distance dominating wavefunctions 
as a function of $\rho$ for the non-natural-parity state 
(i.e. angular momentum conservation forbid the contribution of the 
$^{8}$Be($0^+$) resonance)
$2^-$ (to the left) and the natural-parity state $1^-$ (to the right) 
of $^{12}$C.}
\label{2minus1minus}
\end{figure}

The method allows investigations of a number of the low-lying $^{12}$C resonances, i.e., two
$0^+$, three $2^+$, two $4^+$, and one of each of $1^\pm$, $2^-$, $3^\pm$,
$4^-$ and $6^+$ \cite{alv07}, all below 13.5 MeV.

The resonance structures are changing substantially from small to large
distances. The crucial observables to study the dynamics of the breakup 
process are the momentum distributions of the particles after the decay. 
We use the $2^-$, $4^-$ and $1^-$ resonances of $^{12}$C 
as examples in this report. In our computation we get only
one $2^-$ and one $4^-$ state. We suggest the latter should correspond to the 
state at 6.08 MeV above the triple-$\alpha$
threshold for which a preliminary spin-parity of $2^-$ is assigned in 
\cite{azj}. The 
computed energy distributions could also be used as a tool
to decide this assignment by comparing with measurement. The energies and
widths are compatible with the experimental values. For $1^-$ we get
($E_R$,$\Gamma_R$)=(3.61,0.475) MeV, compared to the experiment 
(3.57,0.315) MeV;
for $2^-$ we get (4.53,0.452) MeV, being the experimental values 
(4.55,0.260) MeV; and for the uncertain $2^-/4^-$ the theoretical values are
(5.98,1.035) MeV and the experimental ones (6.08,0.375) MeV.

\begin{figure}
\vspace*{-2.51cm}
\epsfig{file=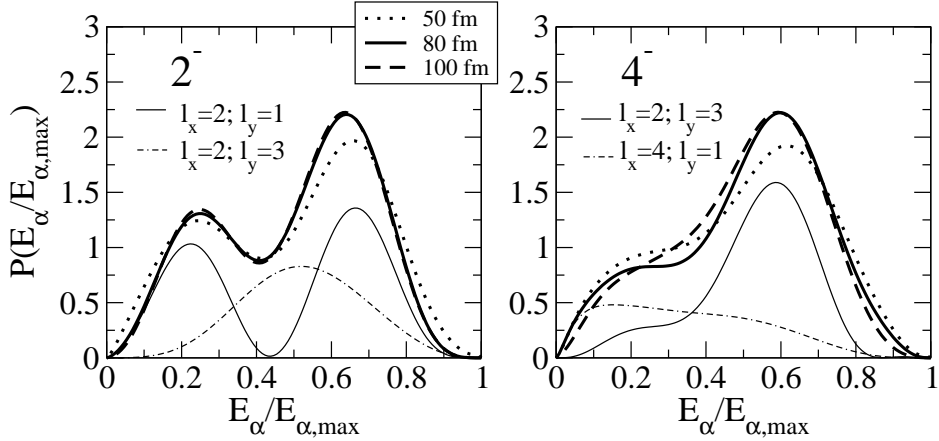,angle=-90,scale=0.51}
\vspace*{-2.3cm}
\caption{To the left, the $\alpha$-particle energy distribution for 
the $2^-$ resonance of $^{12}$C at 11.8 MeV (or 4.53 MeV above the threshold). 
To the right, the $\alpha$-particle energy distribution for the 
$4^-$ resonance of $^{12}$C at 13.3 MeV (or 5.98 
MeV above the threshold).The hyperradii are $\rho = 50,80,100$ fm.
The contribution of the most important partial waves is also shown.
The energy is measured in units of the maximum possible.} 
\label{2minus4minus}
\end{figure}

In ref. \cite{alv07b} we have shown the results corresponding to the 
$0^+$ and $1^+$ states of $^{12}$C. 
In both cases accurate measurements of 
$\alpha$-particle energy distributions are available \cite{dig06}.
The $0^+$ resonances are often approximated by $\alpha$-cluster 
states preferentially decaying sequentially. 
In contrast the $1^+$ resonance has no significant cluster 
structure and is referred to as a shell-model state. Its decay has been
suggested to be direct. Still for both cases the decays into 
three $\alpha$-particles are determined at intermediate and large distances.
The agreement with the accurately measured 
distributions is remarkably good. We conclude that the 
$1^+$ resonance is best described 
by direct decay into the three-body continuum, whereas the $0^+$ 
resonances decay sequentially via $^{8}$Be ground state.

We show in fig.\ref{2minus1minus} the ratios of the four lowest adiabatic 
radial wavefunctions as a function of the hyperradius for the $2^-$ and
$1^-$ resonances located at 4.5 MeV and 3.6 MeV above the triple-$\alpha$ 
threshold, respectively. One can note from the figure that 
only the component relative to the first adiabatic potential contributes 
significantly. This reflects the fact that only
few eigenvalues are needed to get an accurate solution, due to the
fast convergence of the adiabatic expansion. 
The relative sizes of the radial wavefunctions change dramatically around 25 fm
and 15 fm for the $2^-$ and $1^-$ states, respectively. 
The corresponding change of structure around 30 fm
is related to an attempt of populating the 2-body state $^{8}$Be($2^+$).
The ratios of the radial
wavefunctions are also shown to be quite stable under variations of the 
hyperradius at large distances, that is where the energy distributions
are determined.

In fig.\ref{2minus4minus} we plot the $\alpha$-particle energy 
distribution of the $2^-$ and $4^-$ resonances of $^{12}$C for different relatively large
$\rho$-values. The asymptotic behaviour is already reached for hyperradii around 70 fm. 
It can be seen in fig.\ref{2minus4minus} that there is a small variation of
the $4^-$ distribution from 80 to 100 fm. In the $2^-$ distribution 
the curves corresponding to 80 and 100 fm are in fact indistinguishable.
As a test we have Fourier transformed the wavefunction by use of the analytic
solution to the numerically computed adiabatic potentials, which at large
distances is approximated by sums of $1/\rho$ and $1/\rho^2$ terms. 
We obtain very similar distributions with both procedures. Since both
$2^-$ and $4^-$ are non-natural parity states, their decay can not occur 
sequentially via $^8$Be ground state. We are dealing then with a direct
decay into the three-body continuum, or perhaps alternatively with sequential
decay via $^{8}$Be($2^+$). If we estimate the sequential decay via the 
$^{8}$Be($2^+$)
it would produce a peak around $0.3E_{\alpha,max}$ in the $2^-$ and around 
$0.4E_{\alpha,max}$ in the $4^-$ resonance. 
These peaks do not correspond to the ones shown in fig.\ref{2minus4minus},
which means that the 2$^-$ and 4$^-$ energy distributions in the figure
are not consistent with a sequential decay.
The structure in fig.\ref{2minus4minus}
can be understood by separating into individual partial wave 
contributions. The two 
peaks for the $2^-$ state are then related to the two dominating 
terms of $l_x=2$
and $l_y=1,3$, where $l_x$ and $l_y$ are the angular momentum 
between two particles
($l_x$), and their centre of mass and the third particle ($l_y$). 
For the $4^-$-state
both ($l_x$,$l_y$)=(2,3) and (4,1) contribute each with rather 
asymmetric distributions. Thus
we describe these decays as direct into the three-body continuum.

\begin{figure}
\vspace*{-2.51cm}
\epsfig{file=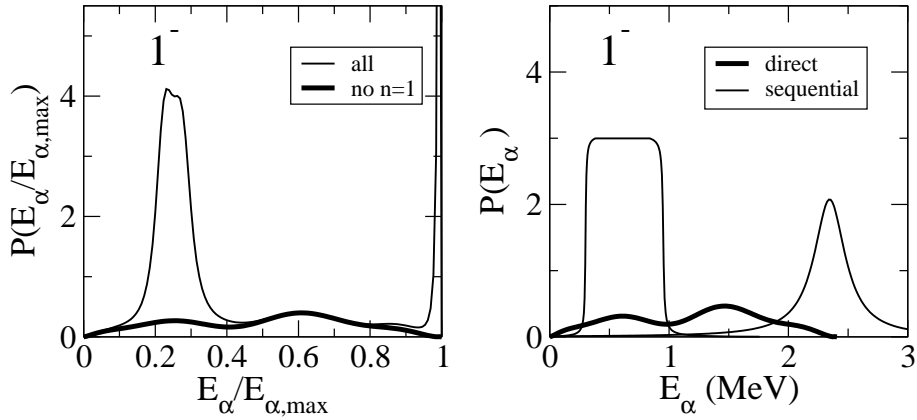,angle=-90,scale=0.51}
\vspace*{-2.3cm}
\caption{The $\alpha$-particle energy distribution for the $1^-$ resonance
of $^{12}$C at 10.89 MeV (or 3.61 MeV above the threshold). 
To the left, the distribution is computed in coordinate space at
$\rho=80$ fm. The thick line correspond to the energy distribution
without the contribution from the first adiabatic component. Its contribution
is about 24\% of the total. To the right, we keep the contribution
without the first adiabatic component, that represents the direct decay.
The sequential decay is illustrated by a Breit-Wigner shape.}
\label{1minus}
\end{figure}

The energy distribution for the $1^-$ resonance is shown in 
fig.\ref{1minus}. The narrow high-energy peak and the distribution around
one quarter of the maximum energy reflect the characteristic features
of a sequential decay via a quasi-stable state. Since the two-body asymptotic
behaviour has not been reached for a hyperradius smaller than 100 fm, the
computed distributions are not accurate, but they provide us with information
about the sequential decay. We use the fact that the sequential decay 
assymptotically must be contained 
in one of the adiabatic potentials, in the present case in the first one. 
We can substitute then the inaccurate
component by the Fourier transform of the known asymptotic two-body behaviour.
This sequential part of the energy distribution
is approximated by the leading order Breit-Wigner shape for the
first emitted $\alpha$-particle. It has a high-energy peak at the most
probable position (2/3 of the resonance energy). The width is the sum of the
three-body decaying resonance width and the width of the intermediate 
two-body resonance. By kinematic conditions we can compute the energy of
the two $\alpha$-particles emerging from $^8$Be, 
that gives rise to the peak at lower energy.
After removing the contribution from the first adiabatic potential, the 
remaining energy distribution is described as direct decay by the computed
three-body continuum coordinate space wavefunction.
From the results in fig.\ref{1minus} we find that the direct decay is about
24\% of the total distribution. 
The distribution is rather uniform but mostly visible between the 
separate peaks of the sequential decay.

\section{Summary and Conclusions.}

We have applied a general new method to compute the particle-energy 
distributions of some of the three-body decaying $^{12}$C many-body 
resonances. We conjectured
that the energy distributions of the decay fragments are insensitive
to the initial many-body structure. The energy distributions are
then determined by the energy and three-body resonance structure
as obtained in a three-$\alpha$ cluster model. These momentum 
distributions are determined by the coordinate space
wavefunctions at large distances.

We have shown the examples of $2^-$, $4^-$ and $1^-$ states of $^{12}$C.
The preliminary spin-parity assignment of the $2^-$ state at 6.08 MeV 
above the triple-$\alpha$ threshold is suggested to be changed to $4^-$.
Both $2^-$ and $4^-$ resonances are best described by direct decay 
into the three-body continuum, whereas the $1^-$ resonance have 
substantial cluster components and decays preferentially via $^{8}$Be 
ground state. The partial wave structure of the resonance at large 
distances is crucial for the energy distribution.

In conclusion, we predict energy distributions of particles emitted 
in three-body decays. 
Both sequential and direct, and both short and long range interactions 
are treated.

\ack
R.A.R. acknowledges support by a post-doctoral fellowship from 
Ministerio de Educaci\'on y Ciencia (Spain).

\end{document}